\documentclass[12pt,english,aps,manuscript]{article}
\usepackage[T1]{fontenc}
\usepackage[latin9]{inputenc}
\usepackage{geometry}
\usepackage{amsmath}
\usepackage{amssymb}
\usepackage{esint}
\usepackage[numbers]{natbib}
\usepackage{babel}
\begin{document}
\begin{center}
\textbf{\Large{}Exact RG Flow Equations and Quantum Gravity}
\par\end{center}{\Large \par}

\begin{center}
\vspace{0.3cm}
 
\par\end{center}

\begin{center}
{\large{}S. P. de Alwis$^{\dagger}$ }
\par\end{center}{\large \par}

\begin{center}
Physics Department, University of Colorado, \\
 Boulder, CO 80309 USA 
\par\end{center}

\begin{center}
\vspace{0.3cm}
 
\par\end{center}

\begin{center}
\textbf{Abstract} 
\par\end{center}

We discuss the different forms of the functional RG equation and their
relation to each other. In particular we suggest a generalized background
field version that is close in spirit to the Polchinski equation as
an alternative to the Wetterich equation to study Weinberg's asymptotic
safety program for defining quantum gravity, and argue that the former
is better suited for this purpose. Using the heat kernel expansion
and proper time regularization we find evidence in support of this
program in agreement with previous work.

\begin{center}
\vspace{0.3cm}
 
\par\end{center}

\vfill{}

$^{\dagger}$ dealwiss@colorado.edu

\eject

\section{Introduction}

Following the seminal work by Ken Wilson \citep{Wilson:1973jj} many
authors have discussed the formulation and consequences of continuum
exact renormalization group (RG) equations for quantum field theory
(QFT). Amongst these the most popular have been those of Polchinski
\citep{Polchinski:1983gv} and Wetterich \citep{Wetterich:1992yh}(see
also \citep{Morris:1993qb,Morris:1994ie,Morris:1994ki})\footnote{For reviews and references to recent work see for example \citep{Morris:1998da}
\citep{Bagnuls:2000ae,Gies:2006wv,Rosten:2010vm}. For applications
to the asymptotic safety program see \citep{Codello:2008vh,Reuter:2012id}\citep{Percacci:2017}
and references therein. }. The former is a differential equation in RG ``time'' $\ln\Lambda$
for the Wilsonian effective action $I_{\Lambda}[\phi]$ obtained by
integrating out the ultra-violet (UV) degrees of freedom down to some
scale $\Lambda$. The latter is a differential equation for the so-called
average effective action, obtained from the functional integral for
the quantum effective action $\Gamma[\phi_{c}]$ by cutting off the
integral over the eigenmodes of the kinetic operator of the QFT at
some infra-red (IR) scale $k$. This produces a functional $\Gamma_{k}[\phi_{c}]$
such that its $k\rightarrow0$ limit gives back $\Gamma[\phi_{c}]$.
It is claimed that this equation defines the evolution all the way
from an ``initial'' UV action all the way down to the deep IR $k\rightarrow0$.

The standard model and Einstein's theory of gravity are usually regarded
as effective field theories (EFT's). The UV completion of these EFT's
is one of the main motivations for string theory. In the latter case
it is expected that these EFT's are valid only up to the string scale,
which is typically an order of magnitude or so below the four dimensional
Planck scale\footnote{If we have a large volume compactification the scale at which the
EFT breaks down is the Kaluza-Klein scale.}. It is commonly believed that above such a UV scale one needs to
replace the EFT by string theory, with the parameters of the EFT being
determined by the fundamental theory through matching conditions in
the transitional region defined by the UV cutoff $\Lambda$.

An alternative to such a situation was proposed four decades ago by
Weinberg \citep{Weinberg:1976xy,Weinberg:1979qg}. He argued that
if the theory of gravity, or indeed gravity coupled to the standard
model, possessed an ultra-violet fixed point with a finite number
of relevant operators, then one would have a finite and predictive
theory at all energy scales. 

Such an eventuality would appear to eliminate the hope that a deeper
understanding of the fundamental laws of nature would explain the
values of the parameters of the standard model such as the Yukawa
couplings, the CKM matrix elements, and the existence of three generations.
On the other hand although it had been hoped that string theory would
provide such an explanation, the discovery that there is an extremely
large (if not infinite) landscape of (semi-realistic) solutions of
the equations of string theory (with different numbers of generations
etc.), means that in practice (at least with our current understanding
of string theory), it is not possible to answer these questions. It
is thus of great interest from a fundamental theoretical point of
view to answer Weinberg's question. 

It is of course well-known that Einstein's theory is not perturbatively
renormalizable and will require an increasing number of counter-terms
if one tries to calculate higher orders in the gravitational constant.
The perturbation series is in effect an expansion in $g(E)=GE^{2}$
where $G$ is the gravitational coupling constant ($\sqrt{8\pi G}=1/M_{P}\sim1/(10^{18}{\rm GeV})$
and $E$ is the energy scale which we are interested in. Thus it is
valid only for energy scales below the Planck scale. Furthermore at
a given energy scale to a given level of accuracy we need only a finite
number of experimentally determined parameters to calculate any scattering
process. However this procedure clearly breaks down as one approaches
the Planck scale and certainly cannot address issues of post-Planckian
physics. In contrast the ``asymptotic safety'' (AS) program initiated
by Weinberg if indeed it can be realized in practice, should be able
to calculate any gravitational process at arbitrarily high energies
in terms of a finite number of parameters.

There is however a practical problem associated with trying to implement
this program - the problem of truncation. The general Wilsonian effective
action valid below some UV scale $\Lambda$ will have an infinite
number of local operators scaled by some (inverse) power of $\Lambda$.
(see for instance equation (\ref{eq:Igravscalar}) below.) In order
to demonstrate the existence of an UV fixed point one needs to compute
the beta functions, and clearly since one is not in the perturbative
regime around the Gaussian fixed point, there is no small parameter
that can give a controlled expansion to approximate the exact $\beta$-function.
Consequently what has been resorted to in the literature has been
a truncation procedure. This has been done in many works (see the
reviews \citep{Codello:2008vh,Reuter:2012id}\citep{Percacci:2017}
and references therein). The procedure may be summarized as follows. 
\begin{itemize}
\item Find a convenient truncation to a subset of the complete set of independent
operators in the action and compute the $\beta$-functions for this
subset. Then find a (non-trivial) fixed point (if one exists) and
compute the scaling exponents there. 
\item Add new couplings and repeat the procedure. 
\item If the enlarged system does not admit a fixed point then clearly the
fixed point of the original truncation was an artifact of the truncation.
If it does then one can compare the fixed point coordinates of the
original set in the two truncations. If these have been preserved
up to small corrections in the enlarged truncation then this may be
interpreted as evidence for a true fixed point of the complete theory.
\end{itemize}
This procedure has been checked against other methods of calculation
in several condensed matter systems - for recent calculations showing
good agreement see for example \citep{Canet:2003qd,Litim:2010tt,Braun:2010tt,Eichhorn:2013zza,Knorr:2016sfs}.

In the gravity case of course there is nothing to compare with and
what has been done is the following. In pure gravity the fixed point
derived from studying the truncation to Einstein gravity and a cosmological
constant term has been shown to be stable under the addition of Riemann
and Ricci squared terms as well as powers of $R$ up to $R^{35}$.
This has been taken as evidence that the fixed point found in the
original truncation is indeed a true fixed point of the theory. Furthermore
numerical investigations (see \citep{Rechenberger:2012pm} and references
therein), have shown the existence of an RG trajectory that connects
such a fixed point with the Gaussian fixed point. This seems to point
towards the existence of a QFT of gravity that accounts for the experimentally
verified weak coupling calculations of Einstein's theory and is asymptotically
complete.

On the other hand, the fact that explicit calculations of the UV fixed
point values of the cosmological and gravitational coupling constants
seem to depend only very weakly on the higher dimension operators,
indicates that some deep principle that is yet to be understood is
underlying this program. If this is uncovered then works such as that
of \citep{Shaposhnikov:2009pv}\citep{Eichhorn:2017ylw}\citep{Gies:2017zwf}
where standard model parameters are evaluated using asymptotic safety
within some truncation ansatz, will indeed be true successes of the
program. In the absence of such an understanding the experimental
agreement of these calculations (for the Higgs and top quark masses)
may still be interpreted as further evidence of the validity of the
truncation in the sense that the higher dimension operators that in
principle could have contributed to the calculation, are in fact negligible
as in the pure gravity case.

Finally let us address the question of observables in quantum gravity.
It is of course well-known that only quantities which are diffeomorphism
invariant are observables in the theory. What the asymptotic safety
program hopes to achieve then is to construct a Wilsonian effective
action, which depends on only a finite number of relevant parameters\footnote{This is clearly enunciated in \citep{Weinberg:2009wa}. }
so that in principle gauge invariant observables such as the S-matrix
can be calculated. The asymptotic safety program can also be of relevance
for computing inflationary observables as has been explained by Weinberg
in \citep{Weinberg:2009wa}.

In this paper we will first review the two main versions of the Wilsonian
exact RG equation and discuss the relation between them\footnote{These equations are in effect RG improved one loop equations. As pointed
out to the author by Joe Polchinski, such an equation was first derived
by Weinberg in \citep{Weinberg:1976xy} (see section 8).}. Then we will argue that the Wetterich version is ill-defined at
the initial point of the evolution i.e. at the UV cutoff. Next we
will derive a background field version of an exact RG equation which
is close in spirit to the Polchinski equation, is well defined at
the UV cutoff, and with proper time regularization, gives a simple
expression for the $\beta$ function\footnote{The equation itself has been written down by other authors by conjecturing
that a RG improved one-loop equation may be exact (see for example
\citep{Bagnuls:2000ae}). In this paper we establish the validity
of this conjecture.}. Furthermore it avoids the problem of having to deal with two sets
of fields, a set of background fields as well as a set of expectation
values of quantum fields (so there are two metrics for instance),
that affects the background field formulation of the Wetterich equation.
In practice much of the work that has been done towards establishing
a fixed point for gravitational theories can actually be taken over
for this RG equation for the Wilsonian action. We then discuss the
beta function equations and the evidence for the existence of an ultra-violet
fixed point with a finite number of relevant directions in a theory
of gravity.

\section{Review and comments\label{sec:Review-and-some}}

Our starting point is the (Euclidean) functional integral for connected
correlation functions $W[J]$, 
\begin{equation}
e^{-W[J]}=\int[d\phi]e^{-I[\phi]-\int\sqrt{g}J.\phi}.\label{eq:W}
\end{equation}
Here $I$ is the ``classical'' action and $J$ is an external (classical)
source. For simplicity we will work with scalar fields, but the expressions
can be extended in the obvious ways to gauge fields, fermion fields,
and metric fields, with appropriate tensor contractions and addition
of gauge fixing terms DeWitt-Fadeev-Popov terms etc., and the replacement
of traces and determinants by supertraces and superdeterminants. We
will also employ a (commonly used) condensed notation so that for
any fields $\phi,\psi$, and a differential operator $K(x,y)\equiv-\nabla^{2}\frac{1}{g^{1/4}(x)}\delta(x-y)\frac{1}{g^{1/4}(y)}$,
\begin{eqnarray*}
\phi.\psi & \equiv & \int\sqrt{g}d^{4}x\phi(x)\psi(x)\\
\phi.K.\psi & \equiv & \int\sqrt{g(x)}d^{4}x\int\sqrt{g(y})d^{4}y\phi(x)K(x,y)\psi(y)
\end{eqnarray*}
Let us now separate the kinetic and interaction parts and write
\[
I[\phi]=I_{0}[\phi]+I_{i}[\phi],
\]
with $I_{0}\equiv\frac{1}{2}\phi.K.\phi]$ is the kinetic term and
$I_{i}$ is the interaction. By standard manipulations (\ref{eq:W})
may be rewritten as,
\begin{equation}
e^{-W[J]}=e^{-I_{i}[-\frac{\delta}{\delta J}]}e^{-\frac{1}{2}{\rm Tr}\ln K+\frac{1}{2}J.K^{-1}.J}\label{eq:W1}
\end{equation}
where $K^{-1}$ is the Green's function associated with $K$. This
expression is formal and (at least in a perturbative expansion in
powers of the interaction) has divergences in any non-trivial field
theory. Thus it needs to be regulated in the ultra-violet (UV). Assuming
that $K$ is a positive operator with eigenvalues $p^{2}$ we impose
a cutoff $\Lambda$ such that the modes $p^{2}\gg\Lambda^{2}$ are
suppressed. A simple hard cutoff would be to replace the Fourier transformed
Green's function $\tilde{K}^{-1}(p^{2})$ by $\tilde{K}_{\Lambda}^{-1}(p^{2})=\tilde{K}^{-1}(p^{2})\theta(\Lambda^{2}-p^{2})$.
In flat space a smoothly cutoff propagator would be for instance of
the form $\tilde{K}_{\Lambda}^{-1}(p^{2})=\tilde{K}^{-1}(p^{2})\exp(-p^{2}/\Lambda^{2})$.
Thus we have
\begin{equation}
e^{-W[J]}=e^{-I_{i\Lambda}[-\frac{\delta}{\delta J}]}e^{-\frac{1}{2}{\rm Tr}\ln K_{\Lambda}+\frac{1}{2}J.K_{\Lambda}^{-1}.J}.\label{eq:W1-1}
\end{equation}
Demanding that the LHS of this equation be independent of $\Lambda$
gives us the following:
\begin{eqnarray}
0 & = & e^{-I_{{\rm i\Lambda}}[-\frac{\delta}{\delta J}]}\left[-\frac{\partial I_{{\rm i}\Lambda}}{\partial\ln\Lambda}[-\frac{\delta}{\delta J}]-\frac{1}{2}\frac{\partial}{\partial\ln\Lambda}{\rm Tr\ln K_{\Lambda}}+\frac{1}{2}J.\frac{d}{d\ln\Lambda}K_{\Lambda}^{-1}.J\right]e^{-\frac{1}{2}{\rm Tr}\ln K_{\Lambda}+\frac{1}{2}J.K_{\Lambda}^{-1}.J}\nonumber \\
 & = & e^{-I_{{\rm i\Lambda}}[-\frac{\delta}{\delta J}]}\int[d\phi]e^{-I_{0\Lambda}[\phi]}\left[-\frac{\partial I_{{\rm i}\Lambda}}{\partial\ln\Lambda}[\phi]-\frac{1}{2}\frac{\partial}{\partial\ln\Lambda}{\rm Tr\ln K_{\Lambda}}+\frac{1}{2}\frac{\delta}{\delta\phi}.\frac{d}{d\ln\Lambda}K_{\Lambda}^{-1}.\frac{\delta}{\delta\phi}\right]e^{-J.\phi}\nonumber \\
 & = & \int[d\phi]e^{-I_{0\Lambda}[\phi]}\left[-\frac{\partial I_{{\rm i}\Lambda}}{\partial\ln\Lambda}[\phi]-\frac{1}{2}\frac{\partial}{\partial\ln\Lambda}{\rm Tr\ln K_{\Lambda}}+\frac{1}{2}\frac{\delta}{\delta\phi}.\frac{d}{d\ln\Lambda}K_{\Lambda}^{-1}.\frac{\delta}{\delta\phi}\right]e^{-[I_{i\Lambda}+J.\phi]}\label{eq:P1}\\
 & = & \int[d\phi]\Biggl\{-\frac{\partial I_{{\rm i}\Lambda}}{\partial\ln\Lambda}[\phi]-\frac{1}{2}\frac{\partial}{\partial\ln\Lambda}{\rm Tr\ln K_{\Lambda}}+\frac{1}{2}\left(\frac{\delta I_{i\Lambda}}{\delta\phi}+J\right).\frac{d}{d\ln\Lambda}K_{\Lambda}^{-1}.\left(\frac{\delta I_{i\Lambda}}{\delta\phi}+J\right)\nonumber \\
 &  & -\frac{1}{2}\frac{\delta}{\delta\phi}.\frac{d}{d\ln\Lambda}K_{\Lambda}^{-1}.\frac{\delta I_{i\Lambda}}{\delta\phi}\Biggr\} e^{-[I_{0\Lambda}+I_{i\Lambda}+J.\phi]}.\label{eq:P0}
\end{eqnarray}
The Polchinski equation results from using a external source current
$J$ which goes to zero at (momentum space) scales well below $\Lambda$.
In this case the $J$ dependent terms in the last equation vanish
and we have the (slightly generalized) Polchinski RG equation,
\begin{equation}
\frac{\partial I_{{\rm i}\Lambda}}{\partial\ln\Lambda}[\phi]=-\frac{1}{2}\frac{\partial}{\partial\ln\Lambda}{\rm Tr\ln K_{\Lambda}}+\frac{1}{2}\left(\frac{\delta I_{i\Lambda}}{\delta\phi}\right).\frac{d}{d\ln\Lambda}K_{\Lambda}^{-1}.\left(\frac{\delta I_{i\Lambda}}{\delta\phi}\right)-\frac{1}{2}\frac{\delta}{\delta\phi}.\frac{d}{d\ln\Lambda}K_{\Lambda}^{-1}.\frac{\delta I_{i\Lambda}}{\delta\phi}.\label{eq:P}
\end{equation}
The slight generalization here is the occurrence of the first term,
which in a general metric background needs to be kept\footnote{This has already been observed in \citep{Jacobson:2012ek}. }
unlike in the original flat space case. Writing $I_{\Lambda}=I_{0\Lambda}+I_{i\Lambda}=I_{0\Lambda}+\sum_{i}g_{\Lambda}^{i}\Phi_{i}[\phi]$
and $\beta^{i}\equiv\frac{\partial g^{i}}{\partial\ln\Lambda}$ gives
from (\ref{eq:P}) as is well known, a set of equations for the beta
functions $\beta_{i}$.

In addition to this UV cutoff we will, in order to connect to the
discussion of the so-called average effective action, introduce an
infra-red (IR) cutoff $k^{2}$. So for example we could have in terms
of a hard cutoff $\tilde{K}_{\Lambda,k}^{-1}(p^{2})=\tilde{K}^{-1}(p^{2})\theta(\Lambda^{2}-p^{2})\theta(p^{2}-k^{2})$,
or in terms of a smooth cutoff $\tilde{K}_{\Lambda,k}^{-1}(p^{2})=\tilde{K}^{-1}(p^{2})[\exp(-p^{2}/\Lambda^{2})-\exp(-p^{2}/k^{2})]$.
In this case (\ref{eq:W1-1}) is replaced by 
\begin{equation}
e^{-W_{k}[J]}=e^{-I_{i\Lambda}[-\frac{\delta}{\delta J}]}e^{-\frac{1}{2}{\rm Tr}\ln K_{\Lambda,k}+\frac{1}{2}J.K_{\Lambda,k}^{-1}.J}.\label{eq:Wk}
\end{equation}
 The dependence on the IR cutoff is indicated by the subscript $k$
on $W$ but it is still independent of the UV scale $\Lambda$. As
discussed above this is equivalent to satisfying the Polchinski equation
(\ref{eq:P}) since the replacement $K_{\Lambda}\rightarrow K_{\Lambda,k}$
has no effect on the derivation of that equation.

Let us now derive the so-called Wetterich equation in a slightly modified
form, i.e. with both a UV cutoff as well as the original IR cutoff.
But now we need to begin with the functional integral version of (\ref{eq:Wk}).
So we have (putting $k\partial_{k}x\equiv\dot{x}$)
\begin{eqnarray*}
\frac{\partial}{\partial\ln k}e^{-W_{k}[J]} & = & -\int[d\phi]\frac{1}{2}\phi.\dot{K}_{k,\Lambda}.\phi e^{-[\frac{1}{2}\phi.K_{k,\Lambda}.\phi+I_{1\Lambda}[\phi]+J.\phi]}\\
 & = & -\frac{1}{2}\frac{\delta}{\delta J}.\dot{K}_{k,\Lambda}.\frac{\delta}{\delta J}e^{-W_{k}[J]}.
\end{eqnarray*}
This leads to the RG equation
\begin{equation}
\dot{W_{k}}|_{J}=\frac{1}{2}\frac{\delta W_{k}}{\delta J}.\dot{K}_{k,\Lambda}.\frac{\delta W_{k}}{\delta J}-\frac{1}{2}\frac{\delta}{\delta J}.\dot{K}_{k,\Lambda}.\frac{\delta W_{k}}{\delta J}.\label{eq:Wflow}
\end{equation}
Defining the cutoff effective action by a Legendre transformation,
\[
\Gamma_{k}[\phi_{c}]=W_{k}[J]-J.\phi_{c},\,\,\phi_{c}=\frac{\delta W_{k}[J]}{\delta J}\Rightarrow\frac{\delta\Gamma_{k}[\phi_{c}]}{\delta\phi_{c}}=-J[\phi_{c}].
\]
Also $\dot{\Gamma}_{k}|_{\phi_{c}}=\dot{W}_{k}|_{J}+\frac{\delta W_{k}[J]}{\delta J}.\dot{J}-\dot{J}.\phi_{c}=\dot{W}_{k}|_{J}$
and $\frac{\delta}{\delta J}.\dot{K}_{k,\Lambda}.\frac{\delta W_{k}}{\delta J}=-{\rm Tr}\dot{K}_{k,\Lambda}\left[\frac{\delta}{\delta\phi_{c}}\otimes\frac{\delta\Gamma_{k}[\phi_{c}]}{\delta\phi_{c}}\right]^{-1}$,
giving from (\ref{eq:Wflow}) a flow equation for the cutoff effective
action,
\[
\dot{\Gamma}_{k}[\phi_{c}]=\frac{1}{2}{\rm Tr}\dot{K}_{k,\Lambda}\left\{ \phi_{c}\otimes\phi_{c}+\left[\frac{\delta}{\delta\phi_{c}}\otimes\frac{\delta\Gamma_{k}[\phi_{c}]}{\delta\phi_{c}}\right]^{-1}\right\} .
\]
The effective average action defined by Wetterich is (except for the
fact that here we are keeping $\Lambda$ finite), 
\[
\Gamma_{k}^{{\rm w}}[\phi_{c}]\equiv\Gamma_{k}[\phi_{c}]-\frac{1}{2}\phi_{c}.R_{k,\Lambda}.\phi_{c},
\]
 where $R_{k,\Lambda}\equiv K_{k,\Lambda}-K_{0,\Lambda}$ (so that
$\dot{R}_{k,\Lambda}=\dot{K}_{k,\Lambda}$). The above equation now
becomes
\begin{equation}
\dot{\Gamma}_{k}^{{\rm w}}[\phi_{c}]=\frac{1}{2}{\rm Tr}\dot{R}_{k,\Lambda}\left[\frac{\delta}{\delta\phi_{c}}\otimes\frac{\delta\Gamma_{k}^{{\rm w}}[\phi_{c}]}{\delta\phi_{c}}+R_{k,\Lambda}\right]^{-1}.\label{eq:Wett}
\end{equation}
In the limit $k\rightarrow0$, $\Gamma_{k}\rightarrow\Gamma$ and
$R_{k,\Lambda}\rightarrow0$, and we have the ``final'' condition
\[
\Gamma_{k}^{{\rm w}}\rightarrow\Gamma,\,{\rm for}\,k\rightarrow0.
\]
To get the initial condition for $\Gamma_{k}^{W}$ let us go back
to the functional integral defining it.
\begin{eqnarray}
e^{-\Gamma_{k}^{{\rm w}}[\phi_{c}]} & = & \int[d\phi]e^{-[\frac{1}{2}\phi.K_{k,\Lambda}.\phi+I_{i\Lambda}[\phi]+J.(\phi-\phi_{c})-\frac{1}{2}\phi_{c}.R_{k,\Lambda}.\phi_{c}]}|_{J=-\delta\Gamma_{k}/\delta\phi_{c}}\nonumber \\
 & = & \int[d\phi']e^{-[\frac{1}{2}\phi_{c}.K_{k,\Lambda}.\phi_{c}+\frac{1}{2}\phi'.K_{k,\Lambda}.\phi'+I_{i\Lambda}[\phi_{c}+\phi']+(J+\phi_{c}.K_{k,\Lambda}).\phi'-\frac{1}{2}\phi_{c}.R_{k,\Lambda}.\phi_{c}]}|_{J=-\delta\Gamma_{k}/\delta\phi_{c}}\nonumber \\
 & = & e^{-[\frac{1}{2}\phi_{c}.K_{0,\Lambda}.\phi_{c}+I_{i\Lambda}[\phi_{c}-\frac{\delta}{\delta J}]}{}^{+\frac{1}{2}{\rm Tr}\ln K_{\Lambda,k}]+\frac{1}{2}(\bar{J}.K_{k,\Lambda}^{-1}.\bar{J})}|_{\bar{J}=-\delta\Gamma_{k}/\delta\phi_{c}+\phi_{c}.K_{k,\Lambda}.}\label{eq:WettInteg}
\end{eqnarray}
When $k\rightarrow\Lambda$, $K_{k,\Lambda}^{-1}\rightarrow0$ and
up to a field independent infinite constant ($\frac{1}{2}{\rm Tr}\ln K_{\Lambda,\Lambda}]$
which can be regularized to zero in heat kernel proper time regularization
for instance), we have 
\[
\Gamma_{k}^{{\rm w}}[\phi_{c}]\rightarrow I_{\Lambda}[\phi_{c}]\,{\rm for}\,k\rightarrow\Lambda.
\]
This establishes the fact that $\Gamma_{k}^{{\rm w}}[\phi_{c}]$ interpolates
between the seed action (i.e. the Wilsonian action at the initial
scale $\Lambda$) and the quantum effective action.

However the flow equation itself breaks down at the initial scale.
This is because as $k\rightarrow\Lambda$, $K_{\Lambda,k}^{-1}\rightarrow0$
and hence $R_{k,\Lambda}=K_{k,\Lambda}-K_{0,\Lambda}\rightarrow\infty$
and so in general does its log derivative. This means that the RHS
of (\ref{eq:Wett}) and hence the RG time derivative of $\Gamma_{k}^{W}$
at the initial time is not well-defined. Thus it is not clear how
the equation can be used to evolve the initial data, i.e. the Wilsonian
action at scale $\Lambda$.

To be concrete let us take the regulated propagator to be 
\[
K_{k,\Lambda}^{-1}(x,y)=<x|\int_{1/\Lambda^{2}}^{1/k^{2}}dse^{-\hat{K}s}|y>.
\]
 Then we have the operator relation (writing $K_{k,\Lambda}^{-1}(x,y)=<x|\hat{K}_{k,\Lambda}^{-1}|y>$
etc.)
\begin{eqnarray}
\hat{R}_{k,\Lambda} & = & \hat{K}_{k,\Lambda}-\hat{K}_{0,\Lambda}=\hat{K}e^{\hat{K}/\Lambda^{2}}(e^{(1/k^{2}-1/\Lambda^{2})\hat{K}}-1)^{-1}\label{eq:Rhat}\\
\dot{\hat{R}}_{k,\Lambda}R_{k,\Lambda}^{-1} & = & -2\frac{\hat{K}}{k^{2}}e^{(1/k^{2}-1/\Lambda^{2})\hat{K}}(e^{(1/k^{2}-1/\Lambda^{2})\hat{K}}-1)^{-1}\label{eq:RdotRinv}
\end{eqnarray}
Both these quantities diverge as $(1/k^{2}-1/\Lambda^{2})^{-1}$ as
$k\rightarrow\Lambda$. Thus the equation (\ref{eq:Wett}) is not
well defined in this limit.

In the literature on the average effective action, in contrast to
the above, there is no explicit UV cutoff. It is implicitly assumed
that one can work with $\Lambda\rightarrow\infty$. In this case (\ref{eq:RdotRinv})
becomes
\[
\dot{\hat{R}}_{k,\infty}R_{k,\infty}^{-1}=-2\frac{\hat{K}}{k^{2}}e^{(1/k^{2})\hat{K}}(e^{(1/k^{2})\hat{K}}-1)^{-1}
\]
It is clear that $k\rightarrow\infty$ is now a well defined limit
with $\dot{\hat{R}}_{k,\infty}R_{k,\infty}^{-1}\rightarrow-2$ . However
this assumes that the initial action $I_{\Lambda}$ is well-defined
in the $\Lambda\rightarrow\infty$ limit which is tantamount to assuming
(at the very least) that the theory has a UV fixed point!

Actually the situation is worse than that since even if a fixed point
existed as in QCD for example, the coupling $g_{\Lambda}\rightarrow0$
as the cutoff $\Lambda\rightarrow\infty$ so that the action $-1/g_{\Lambda}^{2}\int F^{2}$
does not exist\footnote{One may redefine the gauge field to get a canonical kinetic term $A\rightarrow A_{c}=g_{\Lambda}A$,
but in this case the gauge transformation $A_{c}\rightarrow{\cal G}^{-1}A_{c}{\cal G}+\frac{1}{g_{\Lambda}}{\cal G}^{-1}d{\cal G}$
is ill-defined in the limit $\Lambda\rightarrow\infty$. }. One can of course get arbitrarily close to the fixed point with
a well defined action but there is no finite action starting point
as is assumed in the derivations of the Wetterich equation\footnote{For alternative discussions of this so-called reconstruction problem
for the effective average action see \citep{Manrique:2009tj,Vacca:2011fx,Morris:2015oca}. }. One needs to turn on the (marginally) relevant operator above at
some large but finite cutoff $\Lambda$ in order to flow away from
the fixed point as one lowers the cutoff $\Lambda$. Thus even in
this case the assumption that there is a meaningful $\Lambda\rightarrow\infty$
action is invalid. 

The version of the Wetterich equation that is appropriate for gauge
and gravity theories is formulated using the background field method.
However this too has the above problem. Also it is clear from (\ref{eq:WettInteg})
that in the limit $k\rightarrow\Lambda$ there is now a background
field dependent infinity which is absent if regularized for instance
as in eqn (\ref{eq:regulator-1}). Furthermore this formulation requires
the introduction of an effective action which depends on both a background
field and the expectation value of the quantum field - thus it has
two metrics for instance. In the next section we will suggest an alternative
exact RG equation that is well defined at an arbitrary but finite
UV scale $\Lambda$ and is dependent only on the background field.

\section{An alternative RG equation}

The quantum theory corresponding to a given classical action $I[\phi]$
is given by the quantum effective action $\Gamma(\phi_{c})$ defined
(implicitly and formally) by the formula 
\begin{equation}
e^{-\Gamma(\phi_{c})}=\int[d\phi]e^{-I[\phi]-J.(\phi-\phi_{c})}|_{J=-\partial\Gamma/\partial\phi_{c}}.\label{eq:Gamma}
\end{equation}
By translating the integration variable $\phi=\phi_{c}+\phi'$ we
have the following expressions,
\begin{eqnarray}
e^{-\Gamma(\phi_{c})} & = & \int[d\phi']e^{-I[\phi_{c}+\phi']-J.\phi'}|_{J=-\partial\Gamma/\partial\phi_{c}}\nonumber \\
 & = & \int[d\phi']e^{-\{I[\phi_{c}]+\frac{1}{2}\phi'.\frac{\delta^{2}I}{\delta\phi_{c}^{2}}.\phi'+I_{{\rm i}}[\phi_{c},\phi']+(J+\frac{\delta I[\phi_{c}]}{\delta\phi_{c}}).\phi'\}}|_{J=-\delta\Gamma/\delta\phi_{c}}\label{eq:Gamma2}\\
 & = & e^{-I[\phi_{c}]}e^{-\frac{1}{2}{\rm Trln}K[\phi_{c}]}e^{-I_{{\rm i}}[\phi_{c},-\frac{\delta}{\delta\bar{J}}]}e^{\frac{1}{2}\bar{J}.K[\phi_{c}]^{-1}.\bar{J}}|_{\bar{J}=\delta I[\phi_{c}]/\partial\phi_{c}-\delta\Gamma/\delta\phi_{c}}.\nonumber 
\end{eqnarray}
In the second line above $I_{{\rm i}}[\phi_{c},\phi']$ contains all
powers of $\phi'$ which are higher than quadratic in the expansion
of $ $ $I[\phi_{c}+\phi']$, and the third line is the result of
doing the Gaussian integral over $\phi'$. 

The above is a formal expression that needs to be regularized. A convenient
way of doing this for our purposes is to introduce the Schwinger proper
time regularization\footnote{In general $K$ will of course be a matrix over space-time indices
as well as internal indices labelling the different fields as well
as their components. } as in the previous section, except that now the operators depend
on the field $\phi_{c}$. 
\begin{equation}
K_{k,\Lambda}^{-1}(\phi_{c};x,y)=<x|\int_{1/\Lambda^{2}}^{1/k^{2}}dse^{-\hat{K}[\phi_{c}]s}|y>,\,{\rm ln}K_{k,\Lambda}[\phi_{c};x,y]=-<x|\int_{1/\Lambda^{2}}^{1/k^{2}}\frac{ds}{s}e^{-\hat{K}[\phi_{c}]s}|y>.\label{eq:regulator-1}
\end{equation}
So we replace (\ref{eq:Gamma2}) by 
\begin{equation}
e^{-\Gamma_{k}(\phi_{c})}=e^{-I{}_{\Lambda}[\phi_{c}]}e^{-\frac{1}{2}{\rm Trln}K_{k,\Lambda}[\phi_{c}]}e^{-I_{{\rm i}}[\phi_{c},-\frac{\delta}{\delta\bar{J}}]}e^{\frac{1}{2}\bar{J}.K_{k,\Lambda}[\phi_{c}]^{-1}.\bar{J}}|_{\bar{J}=\delta I_{\Lambda}[\phi_{c}]/\partial\phi_{c}-\delta\Gamma_{k}/\delta\phi_{c}}.\label{eq:Gamma3}
\end{equation}
Now we have the following limits:
\begin{equation}
k\rightarrow0,\,\Gamma_{k}\rightarrow\Gamma;\,\,k\rightarrow\Lambda,\,\,K_{k,\Lambda}^{-1},{\rm \,ln}K_{k,\Lambda}\rightarrow0,\,\Rightarrow\Gamma_{k}\rightarrow I_{\Lambda}.\label{eq:limits}
\end{equation}
It should be noted that the problematic limit $k\rightarrow\Lambda$
in the Wetterich case is taken care of by the use of the (regularized)
Schwinger proper time representation for the one loop effective action
- i.e the second equation in (\ref{eq:regulator-1}). The necessity
of the separate regularization of the logarithm of $\hat{K}$ is nothing
but the well known phenomenon that any higher covariant derivative
type regularization (which is essential for the discussion of a gauge
theory) will not regularize the one-loop term. With these regularizations
(\ref{eq:regulator-1})\footnote{One can of course use a whole class of such regularizations with a
smooth cutoff function of the proper time $s$ in the integrands.
Here we just chose the simplest version.} we have the well-defined expression (\ref{eq:Gamma3}).

Differentiating w.r.t. $\ln k$ we have from (\ref{eq:Gamma3}) ($dt\equiv dk/k)$
\begin{eqnarray}
-e^{-\Gamma_{k}(\phi_{c})}\dot{\Gamma}_{k}[\phi_{c}] & = & e^{-I_{\Lambda}[\phi_{c}]}e^{-\frac{1}{2}{\rm Trln}K_{k,\Lambda}[\phi_{c}]}e^{-I_{{\rm i}\Lambda}(\phi_{c},-\frac{\delta}{\delta\bar{J}})}\times\nonumber \\
 &  & (-\frac{1}{2}\frac{d}{dt}{\rm Tr}\ln K_{k,\Lambda}[\phi_{c}]+\dot{\bar{J}}.K_{k,\Lambda}^{-1}.\bar{J}+\frac{1}{2}\bar{J}.\dot{K}_{k,\Lambda}^{-1}.\bar{J})\times\nonumber \\
 &  & e^{\frac{1}{2}\bar{J}.K_{k,\Lambda}[\phi_{c}]^{-1}.\bar{J}}|_{\bar{J}=\delta I_{\Lambda}[\phi_{c}]/\partial\phi_{c}-\delta\Gamma_{k}/\delta\phi_{c}}\label{eq:Gdot}
\end{eqnarray}
Now let us take the limit $k\rightarrow\Lambda$ of this equation.
Using (\ref{eq:limits}) we see that since $I_{{\rm i}}(\phi_{c},-\frac{\delta}{\delta\bar{J}})$
is at least third order in $\delta/\delta J$, it will commute with
the $\bar{J}$ terms up to terms which have at least one power of
$\delta/\delta J$ acting on the last factor which is equal to $1$
in this limit. Also in this limit $\bar{J}\rightarrow0$. Thus we
have the alternate RG equation 
\begin{equation}
\Lambda\frac{d}{d\Lambda}I_{\Lambda}[\phi_{c}]=\frac{1}{2}k\frac{d}{dk}{\rm Tr}\ln K_{k,\Lambda}[\phi_{c}]|_{k=\Lambda}={\rm Tr}\exp\{-\frac{1}{\Lambda^{2}}\frac{\delta}{\delta\phi_{c}}\otimes\frac{\delta}{\delta\phi_{c}}I_{\Lambda}[\phi_{c}]\},\label{eq:P3}
\end{equation}
where in the last step we used (\ref{eq:regulator-1}). The same equation
may be obtained by requiring the independence from $\Lambda$ of $\Gamma_{k}$,
as is required for consistency, in (\ref{eq:Gamma3}) and then taking
the limit $k\rightarrow\Lambda$ . Thus instead of (\ref{eq:Gdot})
we have 
\begin{eqnarray}
0 & = & e^{-I_{\Lambda}[\phi_{c}]}e^{-\frac{1}{2}{\rm Trln}K_{k,\Lambda}[\phi_{c}]}e^{-I_{{\rm i}\Lambda}[\phi_{c},-\frac{\delta}{\delta\bar{J}}]}\times\nonumber \\
 &  & \left\{ -\Lambda\frac{d}{d\Lambda}I_{\Lambda}[\phi_{c}]-\Lambda\frac{d}{d\Lambda}\left(\frac{1}{2}{\rm Tr}\ln K_{k,\Lambda}[\phi_{c}]+I_{{\rm i}\Lambda}[\phi_{c},-\frac{\delta}{\delta\bar{J}}]\right)+\left(\Lambda\frac{d}{d\Lambda}\bar{J}\right).K_{k,\Lambda}^{-1}.\bar{J}+\frac{1}{2}\bar{J}.\left(\Lambda\frac{d}{d\Lambda}K_{k,\Lambda}^{-1}\right).\bar{J}\right\} \times\nonumber \\
 &  & e^{\frac{1}{2}\bar{J}.K_{k,\Lambda}[\phi_{c}]^{-1}.\bar{J}}|_{\bar{J}=\delta I_{\Lambda}[\phi_{c}]/\partial\phi_{c}-\delta\Gamma_{k}/\delta\phi_{c}}\label{eq:Gdot-1}
\end{eqnarray}
By the same argument as before in the limit $k\rightarrow\Lambda$
we get (after using the explicit regulator (\ref{eq:regulator-1})
in the last step) 
\begin{eqnarray*}
\Lambda\frac{d}{d\Lambda}I_{\Lambda}[\phi_{c}] & = & -\frac{1}{2}\Lambda\frac{d}{d\Lambda}{\rm Tr}\ln K_{k,\Lambda}[\phi_{c}]|_{k=\Lambda}={\rm Tr}\lim_{k\rightarrow\Lambda}\int_{1/\Lambda^{2}}^{1/k^{2}}\frac{ds}{s}s\Lambda\frac{dK_{\Lambda}}{d\Lambda}e^{-sK_{\Lambda}[\phi_{c}]}+{\rm Tr}e^{-\frac{1}{\Lambda^{2}}K[\phi_{c}]}\\
 & = & {\rm Tr}\exp\{-\frac{1}{\Lambda^{2}}\frac{\delta}{\delta\phi_{c}}\otimes\frac{\delta}{\delta\phi_{c}}I_{\Lambda}[\phi_{c}]\},
\end{eqnarray*}
which is the same as (\ref{eq:P3}).

The above discussion of course needs to be modified when the set of
fields $\phi$ includes gauge (and graviton) fields. The quadratic
``kinetic'' term in $\phi'$ will have an additional ``gauge fixing''
term which will be also regularized in the same way. So we make the
replacement (with $\alpha$ a gauge fixing parameter) 
\[
K_{k,\Lambda}[\phi_{c}]\rightarrow K_{k,\Lambda}[\phi_{c}]+K_{k,\Lambda}^{{\rm GF}}[\phi_{c},\alpha].
\]
In addition we have the ghost term which is just a determinant term
and will give an additional factor 
\[
e^{+\frac{1}{2}{\rm Trln}K_{k,\Lambda}^{{\rm ghost}}[\phi_{c}]},
\]
on the RHS of (\ref{eq:Gamma3}). Thus the RG equation (\ref{eq:P3})
is replaced by
\begin{eqnarray}
\Lambda\frac{d}{d\Lambda}I_{\Lambda}[\phi_{c}] & = & \frac{1}{2}k\frac{d}{dk}\left\{ {\rm Tr}\ln\left(K_{k,\Lambda}[\phi_{c}]+K_{k,\Lambda}^{{\rm GF}}[\phi_{c},\alpha]\right)-{\rm Tr\ln K_{k,\Lambda}^{ghost}}\right\} |_{k=\Lambda},\nonumber \\
 & = & {\rm Tr}\exp\left\{ -\frac{1}{\Lambda^{2}}\left(I_{\Lambda}^{(2)}[\phi_{c}]+I_{\Lambda}^{(2){\rm GF}}[\phi_{c},\alpha]\right)\right\} -{\rm Tr}\exp\left\{ -\frac{1}{\Lambda^{2}}I_{\Lambda}^{(2){\rm ghost}}[\phi_{c},\alpha]\right\} .\label{eq:P4}
\end{eqnarray}
In the last line we have again used our explicit proper time cutoff.
Also $I_{\Lambda}^{(2){\rm GF}}[\phi_{c},\alpha],\,I_{\Lambda}^{(2){\rm ghost}}[\phi_{c},\alpha]$
are the background dependent operators defining the gauge fixing and
ghost terms (which are respectively quadratic in the quantum field
$\phi'$ and the ghost fields $C,\bar{C}$). Also $I_{\Lambda}$ is
as before the Wilsonian action but now including also the gauge fixing
and ghost terms with $\Lambda$ dependent couplings.

We should point out at this stage that this exact RG equation is in
fact simply an RG improvement of the proper time representation of
the RG equation for the cutoff one-loop effective action. As a one
loop equation it has been written down by many authors for example
\citep{Floreanini:1995aj}\citep{Bonanno:2005bj,Bonanno:2004sy}.
In a form that is the same as what we have above it is given in \citep{Bagnuls:2000ae}
(see equation (92)). The point of the above discussion was however
to derive it as an exact equation for the Wilsonian effective action.

The action can be expanded in a complete set of local operators (after
eliminating redundancies) 
\begin{equation}
I_{\Lambda}=\sum\bar{g}^{A}(\Lambda)\Phi_{A}[\phi].\label{eq:Iexapansion}
\end{equation}
The equation (\ref{eq:P3}) is then in effect an RG improved one loop
equation for the beta functions. It is (as is the case for the original
form of the Polchinski equation) an exact equation for the evolution
of the Wilson effective action. It is also the appropriate form to
use for exploring the fixed points (if any) for gauge and gravitational
field theories since it has manifest gauge invariance under the gauge
transformation of $\phi_{c}$. 

In fact all one needs for the exploration of UV fixed points is the
above equation. Indeed what has been done in the literature is completely
equivalent to doing an operator truncation of this equation, since
in practice the equation for the effective average action \citep{Wetterich:1992yh,Reuter:1996cp}
can be used only in a regime where the derivative expansion is valid.
Thus in effect one is dealing with a derivative expansion in terms
of local operators. In this sense $\Gamma_{k}$ is not in any way
different from the Wilsonian effective action. The validity of the
derivative expansion requires that $\partial^{2}/k^{2}\ll1$ i.e.
one cannot really take the $k\rightarrow0$ limit. There is no sense
in which one can get the non-local quantum effective action from the
quasi-local $\Gamma_{k}$ without being able to sum the relevant infinite
series. However the former is not needed to get the RG equations or
to establish the existence of a UV fixed point. In other words one
does not need to have an action which interpolates from the initial
Wilsonian action all the way to the quantum effective action $\Gamma[\phi_{c}]$
to explore the UV properties of the theory. Finally let us emphasize
that this formulation avoids the problem of having two background
fields such as having two metrics - which is the case for the Wetterich
equation.

\subsection{Gauge fixing dependence and observables in quantum gravity\label{sub:Gauge-fixing-dependence}}

DeWitt \citep{DeWitt:2003pm} has argued that the gauge fixing parameter
is not renormalized. If this is the case we may assume that the gauge
fixing parameters do not flow. Furthermore the 1PI effective action
$\Gamma$ at its extremum is in fact independent of the gauge fixing.
On the other hand the Wilsonian action $I_{\Lambda}$ and the average
effective action $\Gamma_{k}$ are gauge fixing dependent\footnote{For a detailed discussion of this and a suggestion for a modified
version of $\Gamma_{k}$ see \citep{Lavrov:2012xz,Lavrov:2015kta}.} and therefore so are the beta functions. Nevertheless the fixed points
and the critical exponents should be gauge independent. 

Currently what has been done is to check the gauge dependence by comparing
the calculations in different gauges and to some extent this has been
verified - see for example the discussion in section 7.3.3 of \citep{Percacci:2017}.
However clearly one needs a general argument establishing this.

One possibility would be to show that the different gauge choices
are equivalent to reparametrizations in field space. An alternative
approach to using gauge fixing is perhaps to use a cutoff version
of the Vilkovisky-DeWitt gauge independent formulation of gauge theory
\citep{DeWitt:2003pm}. We will leave further discussion of this problem
to future work.

The question of gauge dependence also leads us to address what it
is that a theory of quantum gravity hopes to calculate. As is well-known
diffeomorphism invariance in quantum gravity implies that there are
no local observables. In asymptotically flat or AdS backgrounds however
it is possible to define an S-matrix that is well-defined as an observable.
It may even be possible to do so in an asymptotically dS background
\citep{Witten:2001kn}. Thus one would adopt the following procedure. 

Consider the Wilsonian effective action for a theory of quantum gravity
with a cutoff scale $\Lambda$. This would be an expansion in terms
of (space-time integrals of) an infinite set of local operators which
is valid for energy scales $E^{2}\sim\partial\phi$ or $E^{2}\sim R$
(where $\phi$ is any field and $R$ is space-time curvature), such
that $E<\Lambda$, so that the expansion in terms of local operators
is valid. Thus for the pure gravity case we have,
\begin{eqnarray}
I_{\Lambda} & = & \int d^{4}x\sqrt{g}[\Lambda^{4}g_{0}(\Lambda)+\Lambda^{2}g_{1}(\Lambda)R+(g_{2a}(\Lambda)R_{\mu\nu}R^{\mu\nu}+g_{2b}(\Lambda)R^{2}+g_{3b}R_{....}R^{....})\nonumber \\
 &  & +\Lambda^{-2}(g_{3a}(\Lambda)RR_{\mu\nu}R^{\mu\nu}+\ldots)+(g_{3a}^{(1)}R\square R+\ldots)+O(\Lambda^{-4})]+I_{\Lambda}^{{\rm GF}}+I_{\Lambda}^{{\rm ghost}}.\label{eq:Igrav}
\end{eqnarray}
The asymptotic safety program hopes to establish that the dimensionless
couplings have, in addition to the Gaussian fixed point $g_{i}=0$
in the IR, also a non-trivial fixed point $g_{i}=g_{i}^{*},\,\beta_{i}(g^{*})=0$
where not all $g_{i}^{*}$ are zero. Furthermore only a finite number
of these dimensionless couplings $g_{i}$ are expected to be relevant.
For instance in concrete calculations it appears that only $g_{0},g_{1},g_{2}$
are relevant, and so need to be determined by experiment. This then
gives us a Wilsonian effective action which may be used to calculate
the S-matrix for gravitons, once the background for the far past and
the far future of the scattering process has been chosen to be (say)
flat space, for arbitrarily high energies $E<\Lambda$ in terms of
the three undetermined couplings\footnote{A concrete method of calculation (in the particular case of cosmology
at least) has been discussed by Weinberg \citep{Weinberg:2009wa}.
For a given energy $E$ one needs to optimize the value of the cutoff
$\Lambda$. In order to be able to ignore higher dimension operators
one needs $E<\Lambda$. On the other hand in order to be able to ignore
higher order radiative corrections the cutoff should not be much higher
than $E$. This is of course very much in the spirit of perturbative
QCD calculations as explained in the above reference. }. Furthermore the BRST invariance of (\ref{eq:Igrav}) ensures that
the S-matrix is independent of the gauge fixing\footnote{See for instance \citep{DeWitt:2003pm} and also \citep{Antoniadis:1986tu,Tomboulis:2015esa}.}.

One issue that may affect this argument is the question of unitarity.
Perturbatively it appears that any higher derivative theory has a
propagating ghost - in particular in higher derivative gravity there
appears to be a spin two ghost \citep{Stelle:1976gc}. However this
weak coupling argument may not be valid in the complete theory. Let
us consider this in more detail. In the theory defined above (\ref{eq:Igrav})
each dimensionless coupling will have an asymptotic expansion for
large as well as small $\Lambda.$ Thus defining the planck scale
$M_{P}$ by (the inverse of) the gravitational constant measured at
long distances we have for instance with $g_{i}^{(j)}$ being pure
numbers or (for $j>0)$ at most polynomials in $\ln\frac{\Lambda}{M_{P}}$
,
\begin{eqnarray*}
g_{1}(\Lambda) & = & g_{1}^{(0)}+g_{1}^{(1)}\frac{M_{P}^{2}}{\Lambda^{2}}+\ldots,\,\Lambda\gg M_{P}\\
 & = & \frac{M_{P}^{2}}{2\Lambda^{2}}+\tilde{g}_{1}^{(1)}+\tilde{g}_{1}^{(2)}\frac{\Lambda^{2}}{M_{P}^{2}}+\ldots,\,\Lambda\ll M_{P}\\
g_{2}(\Lambda) & = & g_{2}^{(0)}+g_{2}^{(1)}\frac{M_{P}^{2}}{\Lambda^{2}}+\ldots,\,\Lambda\gg M_{P}\\
 & = & \tilde{g}_{2}^{(0)}+\tilde{g}_{2}^{(1)}\frac{\Lambda^{2}}{M_{P}^{2}}+\ldots,\,\Lambda\ll M_{P}.
\end{eqnarray*}
Now the existence of the perturbative ghost is inferred by looking
at the propagator of the low energy theory (in a flat background i.e.
writing $g_{\mu\nu}=\eta_{\mu\nu}+\frac{2}{M_{P}}h_{\mu\nu}$ truncated
to four derivatives), which turns out to have two poles, one at zero
mass corresponding to the graviton and another with squared mass $M_{P}^{2}/\tilde{g}_{2}^{(0)}$.
But at this point the theory is strongly coupled and all higher derivative
terms would also contribute and it is not at all clear that this putative
ghost will survive in the full theory. Indeed it has been argued that
such states decouple and that the S-matrix is unitary \citep{Antoniadis:1986tu}.
Furthermore it has been shown in toy models that the usual argument
(even in weak coupling) for the existence of a ghost in quartic derivative
theories is incorrect \citep{Bender:2007wu,Bender:2008vh,Mannheim:2018ljq}.
The Hamiltonian of the theory has to be interpreted not as a (Dirac)
Hermitian one but as a PT symmetric one - in which case contrary to
the naive expectation one has a unitary theory with a positive energy
spectrum. 

While the above arguments clearly do not imply that (\ref{eq:Igrav})
defines a unitary quantum gravity, what they do show is that the naive
argument for the existence of a perturbative ghost, does not mean
that the correctly interpreted complete theory violates unitarity.

\section{The beta function equations}

\subsection{General considerations}

The formula (\ref{eq:P4}) gives a straightforward way of evaluating
the beta functions of any theory. In (\ref{eq:Iexapansion}) let us
introduce dimensionless couplings $g^{A}$ by writing 
\begin{equation}
\bar{g}^{A}=\Lambda^{4-n_{A}}g^{A}\label{eq:gbarg}
\end{equation}
 where $n_{A}$ is the physical (a.k.a. canonical or engineering)
dimension of the operator $\Phi$. Thus $n_{A}=0$ for the unit operator
(the cosmological constant term proportional to $\sqrt{g}$), $n_{A}=2$
for the Einstein-Hilbert term and for a scalar mass term, $n_{A}=4$
for scalar, vector and fermionic kinetic terms and ``renormalizable''
interactions in the sense of perturbative QFT. Terms with physical
dimensions $n_{A}>4$ are the so-called ``non-renormalizable'' terms,
amongst which one will have both higher derivative terms such as $\phi\square^{2}\phi,\,R^{2},$
as well as higher powers of field operators such as $\phi^{6},(\bar{\psi}\psi)^{2}$.
Then the flow equation (\ref{eq:P3}) (we ignore the complications
of gauge fixing and ghosts for the moment), becomes an infinite set
of coupled equations for the dimensionless couplings $g_{A}$:
\begin{equation}
\dot{g}^{A}+(4-n_{A})g^{A}=\Lambda^{n_{A}-4}{\rm Tr}\exp\{-\frac{1}{\Lambda^{2}}\frac{\delta}{\delta\phi_{c}}\otimes\frac{\delta}{\delta\phi_{c}}I_{\Lambda}[\phi_{c}]\}|_{\Phi_{A}},\,A=0,1,2,\ldots.\label{eq:beta1}
\end{equation}
The instruction on the right is to isolate the coefficient of the
operator $\Phi_{A}$ in the expansion of the trace. Also we've written
$\dot{g}\equiv\frac{dg}{dt},\,t=\ln\frac{\Lambda}{\Lambda_{0}}$ where
$\Lambda_{0}$ is a fiducial scale which can be identified with the
Planck scale (i.e. $\Lambda_{0}^{2}=M_{P}^{2}\equiv1/8\pi G_{N}$
where $G_{N}$ is Newton's constant measured at low energies).

The RHS of (\ref{eq:beta1}) gives the contribution of quantum fluctuations
to the various beta functions. It is succinctly given in this formula
by the heat kernel trace whose expansion in powers of $1/\Lambda^{2}$
can be systematically worked out\footnote{In practice beyond the first few orders, it becomes extremely complicated
though.}. 

These beta function equations take the general form

\begin{equation}
\dot{g}^{A}+(4-n_{A})g^{A}=\eta^{A}(\{g\})\label{eq:beta2}
\end{equation}
with $\eta^{A}=0,\forall A$ when $g^{A}=0,\forall A$ provided we
replace $g^{A}\rightarrow\hat{g}^{A}\equiv(g^{A})^{-1}$ for gravitational
and gauge couplings. This means that there is always a Gaussian fixed
point solution $\dot{g}^{A}=0,\,g^{A}=0,\,\forall A$. A non-trivial
fixed point would exist if the infinite set of equations
\begin{equation}
(4-n_{A})g^{A}=\eta^{A}(\{g\}),\label{eq:NGFP}
\end{equation}
 has real solutions $g^{A}=g_{*}^{A},$ with $g_{A}^{*}$ finite for
all $A$ and $\ne0$ for at least some couplings. We will argue below
using the general structure of the heat kernel expansion, that this
is indeed the case for gravity coupled to a scalar field theory.

The question of the nature of the fixed point and in particular its
critical surface is determined by linearizing (\ref{eq:beta2}) around
the fixed point so we have
\begin{equation}
\frac{d\delta g}{dt}^{A}=\sum_{B}\left(-(4-n_{A})\delta_{B}^{A}+\frac{\partial\eta^{A}(\{g\})}{\partial g^{B}})|_{g_{_{*}}}\right)\delta g^{B}\equiv\sum_{B}D_{\,\,B}^{A}(g_{*})\delta g^{B}.\label{eq:deltagdot}
\end{equation}
The number of negative eigenvalues of the matrix ${\bf D}$ (i.e.
the number of relevant directions) is then the dimensionality of the
critical surface for an UV fixed point. A predictive (i.e. renormalizable)
theory should have only a finite number of relevant directions and
the corresponding couplings (at some fiducial scale) would need to
be determined by experiment. The other (irrelevant) directions can
then be set to their fixed point values. 

To be more specific let us introduce the eigenvectors ${\bf u}$ of
the matrix ${\bf D}$ with eigenvalues $\theta^{(J)}$ - i.e. ${\bf D}{\bf u}^{(J)}=\theta^{(J)}{\bf u}$.
Also suppose that $\theta^{(0)},\ldots\theta^{(R-1)}<0$, while the
rest are positive (or zero). In this case 
\[
{\bf u}^{(J)}(t)=e^{-|\theta^{(J)}|t}{\bf u}^{(J)}(0)\rightarrow0,\,J=0,\ldots R-1,\,t\rightarrow\infty.
\]
 So the deviation from the fixed point value at the fiducial scale
$\Lambda=\Lambda_{0}$ will need to be fixed by experiment. On the
other hand for (the infinite set of) positive (or zero) eigenvalues
we set ${\bf u}^{(J)}(0)=0$, i.e. the corresponding couplings at
the fiducial scale are equal to the fixed point values.

Now canonically in the standard model coupled to gravity there are
only three operators with $n_{A}<4$, namely the unit operator (cosmological
constant term) with $n=0$, the Higgs mass term with $n=2$ and the
Einstein-Hilbert term with $n=2$. Then there are kinetic terms for
all fields and the Yukawa couplings, with canonical dimension $n=4$.
All the other (infinite number of) operators have integral canonical
dimensions with $n\geq5$, i.e. $4-n<-1$. Thus unless there are large
anomalous dimensions (at the NGFP) one might expect the number of
relevant operators to remain the same (or at least finite) as in the
absence of quantum corrections. Obviously this is the case around
the Gaussian fixed point. We will argue below that this is very likely
to be the case around the UV fixed point as well.

\subsection{The heat kernel expansion and the beta functions}

In this section we will discuss the beta functions for a scalar theory
coupled to gravity\footnote{For previous treatments of this system based on the Wetterich equation
see for example \citep{Percacci:2015wwa} and references therein. }. The action for the theory at some scale $\Lambda$ is 
\begin{eqnarray}
I_{\Lambda} & = & \int d^{4}x\sqrt{g}[\Lambda^{4}g_{0}(\Lambda)+\Lambda^{2}g_{1}(\Lambda)R+(g_{2a}(\Lambda)R_{\mu\nu}R^{\mu\nu}+g_{2b}(\Lambda)R^{2}+g_{3b}R_{....}R^{....})\nonumber \\
 &  & +\Lambda^{-2}(g_{3a}(\Lambda)RR_{\mu\nu}R^{\mu\nu}+\ldots)+(g_{3a}^{(1)}R\square R+\ldots)+O(\Lambda^{-4})]\nonumber \\
 &  & +\int d^{4}x\sqrt{g}[Z(\phi^{2}/\Lambda^{2})\frac{1}{2}\phi(-\square)\phi+V(\phi,\Lambda)+\xi(\phi,\Lambda)R+O(\partial^{4})]\nonumber \\
 &  & +I_{\Lambda}^{({\rm G.F}.)}+I_{\Lambda}^{({\rm ghost)}}.\label{eq:Igravscalar}
\end{eqnarray}
Here 
\begin{eqnarray}
V(\phi,\Lambda) & = & \frac{1}{2}\lambda_{1}(\Lambda)\Lambda^{2}\phi^{2}+\frac{1}{4!}\lambda_{2}(\Lambda)\phi^{4}+\frac{1}{6!}\lambda_{3}(\Lambda)\Lambda^{-2}\phi^{6}+\ldots,\label{eq:Vexpansion}\\
Z\left(\frac{\phi^{2}}{\Lambda^{2}}\right) & = & Z_{0}+\frac{1}{2}Z_{1}\frac{\phi^{2}}{\Lambda^{2}}+\ldots\label{eq:Zexpansion}\\
\xi\left(\phi,\Lambda\right) & = & \frac{1}{2}\xi_{1}\phi^{2}+\frac{1}{4!}\xi_{2}\phi^{4}+\ldots\label{eq:xiexpansion}
\end{eqnarray}
 All coefficients $g_{i},\lambda_{i},$ etc. are dimensionless. The
field independent term in the potential has been included in the first
line as an explicit cosmological constant term i.e. $\Lambda^{4}g_{0}(\Lambda)=\Lambda_{CC}$.
Also $\Lambda^{2}g_{1}(\Lambda)=-1/16\pi G_{N}(\Lambda)\equiv-1/2\kappa^{2}(\Lambda)$
where $G_{N}(\Lambda)$ is Newton's constant at the scale $\Lambda$.
To get the matrix $\frac{\delta}{\delta\phi_{c}}\otimes\frac{\delta}{\delta\phi_{c}}I_{\Lambda}[\phi_{c}]\equiv{\bf I}_{\Lambda}^{(2)}$
we expand around the background fields $g_{\mu\nu}\rightarrow g_{\mu\nu}+2\kappa h_{\mu\nu},$
$\phi\rightarrow\phi+\hat{\phi}/\sqrt{Z_{0}}$ (dropping the subscript
$c$), and identify the coefficients of $h\otimes h,h\otimes\hat{\phi},\hat{\phi}\otimes\hat{\phi}$,
to evaluate the second derivative matrix on the background fields.
If one restricts the discussion to the two derivative action then
this matrix operator takes the form 
\begin{equation}
{\bf I}_{\Lambda}^{(2)}=-\nabla^{2}{\bf I}+{\bf E}\label{eq:I2}
\end{equation}
 where ${\bf I}$ is the unit matrix on the space of symmetric transverse
traceless tensors, vectors and scalars as well as space-time, and
${\bf E}$ is a matrix on the same space with matrix elements that
are linear in the Riemann tensor, as well as $\phi$ dependent terms.
The RG flow equation (\ref{eq:P4}) will however generate the higher
derivative terms in the action. Then in addition there will be terms
$\nabla^{4},$ etc in (\ref{eq:I2}) and ${\bf E}$ will have higher
dimension (i.e. greater than or equal to four) field dependent terms;
higher powers of the Riemann tensor as well as its derivatives, in
addition to higher dimension operators constructed out of $\phi$
and its derivatives and mixed terms such as $R\phi^{2}$ etc.

It is convenient to separate the constant part of the matrix ${\bf E}$
by writing 
\begin{equation}
{\bf E}=\Lambda^{2}{\bf E}_{0}+\hat{{\bf E}}.\label{eq:E0Ehat}
\end{equation}
 Here the first constant term comes from the cosmological constant
term (the first term of equation (\ref{eq:Igravscalar})) and the
scalar mass term, so that (labelling the rows and columns of the matrix
schematically with $h,\hat{\phi}$,
\[
{\bf {\bf E}}_{0}=\begin{bmatrix}\frac{g_{0}}{g_{1}}{\bf I} & 0\\
0 & Z_{0}^{-1}\lambda_{1}(\Lambda)
\end{bmatrix}.
\]

The field dependent operator on the other hand has the structure
\[
\hat{{\bf E}}=\begin{bmatrix}\hat{O}_{2} & \hat{O}_{1}\\
\hat{O}_{1}^{{\rm T}} & \hat{\tilde{O}}_{2}
\end{bmatrix},
\]
where the subscripts on the operators indicate the lowest (operator)
dimension contained therein and we have suppressed the matrix indices.
The off diagonal terms come from mixed derivatives such as $\delta^{2}/\delta g\delta\phi$
and are odd dimensional, starting with the dimension one operator
which is linear in $\phi$. Thus it would contain terms such as $g_{1}^{-1/2}Z_{0}^{-1/2}(\lambda_{1}(\Lambda)\Lambda\phi+\frac{1}{3!}\lambda_{2}(\Lambda)\Lambda^{-1}\phi^{3}+\ldots)$
and $\square\phi/(\sqrt{g_{1}Z_{0}}\Lambda)$ , etc. They will contribute
to the traces of quadratic and higher powers of $\hat{{\bf E}}$ in
the heat kernel expansion. The diagonal operators are even dimensional
- starting with terms such as $R$ and $g_{1}^{-1}\Lambda^{-2}V(\phi)$
in the case of the $hh$ block $\hat{O}_{2}$, and with $\frac{1}{2}\lambda_{2}\phi^{2}+\frac{1}{4!}\lambda_{3}\Lambda^{-2}\phi^{4}+\ldots$and
$2\xi(0)R+\ldots$, for the $\phi\phi$ block. In addition there is
a ghost sector which we have suppressed for the moment.

To be explicit let us write out $\hat{{\bf E}}$ in the above theory
of gravity coupled to a scalar field keeping only up to dimension
two operators. Let us also label the rows and columns by $h_{\mu\nu}^{TF},\,h\equiv ih_{\lambda}^{\lambda},\,C_{\mu},\,\hat{\phi}\equiv\sqrt{Z_{0}}\delta\phi$
where the penultimate field is the diffeomorphism ghost, and the $i$
in the trace comes from the rotation of the integration over the conformal
mode to the imaginary axis to get a well-defined Euclidean functional
integral. So including the gauge fixing term in Landau gauge (see
\citep{Reuter:1996cp,Codello:2008vh}) we have 
\[
\hat{O}_{2}=2{\bf U[}({\bf I}-{\bf P})-{\bf P}]+\kappa^{2}\hat{{\bf T},}
\]
where 
\begin{eqnarray}
U_{\rho\sigma}^{\mu\nu} & = & \frac{1}{2}R\delta_{\rho\sigma}^{\mu\nu}+\frac{1}{2}(g^{\mu\nu}R_{\rho\sigma}+g_{\rho\sigma}R^{\mu\nu})-\delta_{(\rho}^{(\mu}R_{\sigma)}^{\nu)}-R_{\,\,(\rho\,\,\sigma)}^{(\mu\,\,\nu)},\label{eq:U}\\
\hat{T}_{\mu\nu}^{\lambda\sigma} & = & \frac{1}{2}(g_{\mu\nu}T^{\lambda\sigma}+g^{\lambda\sigma}T_{\mu\nu})+(\frac{\delta T_{\mu\nu}}{\delta g_{\lambda\sigma}}+\frac{\delta T^{\lambda\sigma}}{\delta g^{\mu\nu}}).\label{eq:T}
\end{eqnarray}
Here $T_{\mu\nu}$ is the stress-energy tensor of the matter sector
and $\delta_{\rho\sigma}^{\mu\nu}=\delta_{\rho}^{(\mu}\delta_{\sigma}^{\nu)}$
($X_{(\mu\nu)}\equiv\frac{1}{2}(X_{\mu\nu}+X_{\nu\mu})$) and ${\bf I}-{\bf P},\,{\bf P}=[\frac{1}{4}g^{\mu\nu}g_{\lambda\sigma}]$
are projection operators onto the space of trace free and traced two
index symmetric tensors. Thus $\hat{O}_{2}$ is a $10\times10$ matrix
acting on the space of symmetric tensors which can be partially diagonalized
into a $9\times9$ matrix on symmetric trace free tensors $h_{\mu\nu}^{(TF)}$
and a $1\times1$ acting on the trace part $h_{\mu}^{\mu}$. Including
also the ghosts, the second diagonal block (labelled by $C^{\mu},\hat{\phi}$)
is (with ${\bf R}=[R_{\mu}^{\,\,\nu}]$)
\[
\hat{\tilde{O}}_{2}=\begin{bmatrix}{\bf R} & {\bf 0}\\
{\bf 0} & \hat{V}^{''}(\phi)+\hat{\xi}^{''}(\phi)R
\end{bmatrix},
\]
where $\hat{V}^{''}(\phi)=\frac{1}{2}\hat{\lambda}_{2}\phi^{2}+\ldots$
and $\hat{\xi}^{''}(\phi)=\hat{\xi}_{1}+\frac{1}{2}Z_{0}\hat{\xi}_{2}\phi^{2}+\ldots$with
$\hat{\lambda}_{i}\equiv\lambda_{i}/(Z_{0})^{i}$ and $\hat{\xi}_{i}\equiv\xi_{i}/(Z_{0})^{i}$
. The off-diagonal blocks take the form
\[
\hat{O}_{1}=\begin{bmatrix}{\bf 0} & \frac{\kappa}{\sqrt{Z_{0}}}{\bf T},_{\phi}\\
0 & i\frac{\kappa}{\sqrt{Z_{0}}}T_{\mu}^{\mu},_{\phi}
\end{bmatrix}
\]
where the row labels have been separated to correspond to the trace
free and trace parts of $h_{\mu\nu}$. The $i$ is a consequence of
the rotation of the trace part (the conformal mode) of the graviton
fluctuation to the imaginary axis.

The heat kernel coefficients for the first few terms of its expansion
have been calculated (see for example \citep{Vassilevich:2003xt}).
As we said earlier it is useful to separate the constant piece so
we have 
\begin{equation}
\exp[-\Lambda^{-2}(-\nabla^{2}{\bf I}+{\bf E})]=e^{-{\bf E}_{0}}\exp[-\Lambda^{-2}(-\nabla^{2}{\bf I}+\hat{{\bf E})}].\label{eq:expK}
\end{equation}
Our object is to calculate the trace of this operator over space-time
and internal indices i.e. ${\rm Tr=\int d^{4}x\sqrt{g}{\rm tr}}$.
Writing ${\bf K}\equiv-\nabla^{2}{\bf I}+{\bf E}$ we have the expansion,
\begin{eqnarray}
{\rm Tr}e^{-{\bf K}/\Lambda^{2}} & = & \frac{1}{(4\pi)^{2}}[\Lambda^{4}B_{0}({\bf K})+\Lambda^{2}B_{2}({\bf K})+B_{4}({\bf K})+\Lambda^{-2}B_{6}({\bf K})+\ldots],\label{eq:Heatexpansion}\\
B_{n}({\bf K}) & = & \int d^{4}x\sqrt{g}{\rm tr}e^{-{\bf E}_{0}}{\bf b}_{n}.
\end{eqnarray}
Let us quote the first three $b$ coefficients:

\begin{eqnarray}
{\bf b}_{0} & = & {\bf I},\label{eq:b0}\\
{\bf b}_{2} & = & \frac{R}{6}{\bf I}-\hat{{\bf E}},\label{eq:b2}\\
{\bf b}_{4} & = & \frac{1}{180}(R^{....}R_{....}-R^{..}R_{..}+\frac{5}{2}R^{2}+6\nabla^{2}R){\bf I}\nonumber \\
 &  & -\frac{1}{6}R\hat{{\bf E}}+\frac{1}{2}\hat{{\bf E}}^{2}-\frac{1}{6}\nabla^{2}\hat{{\bf E}}.\label{eq:b4}
\end{eqnarray}
These coefficients get rapidly more complicated. However for our purposes
the exact numbers in front of the operators are not important. The
only assumption we make is that they are $O(1)$ or less as is the
case for the ones that have been calculated. In the general situation
where one needs to keep higher derivative terms such as $R^{2}$ and
$\phi\square^{2}\phi$ which lead to fourth order derivative operators
in (\ref{eq:I2}), the coefficients of the fourth order operators
(i.e. in ${\bf b}_{2r}$ for $r\ge2$) will of course change, but
will also affect the dimension 0 and 2 operators. 

On the other hand for the scalar potential an exact RG equation is
possible except for the constant (cosmological constant) term, which
depends on all higher derivative couplings. The reason is that in
computing the evolution of the potential one can treat the fields
as constants so that the entire contibution will factor out as in
(\ref{eq:expK}) with ${\bf E}_{0}$ now containing all non-derivative
terms (such as $\phi^{2n}$), and $\hat{{\bf E}}$ now having at least
two derivatives of the fields \footnote{It should be noted that  in spite of the imaginary off-diagonal elements
in ${\bf E}$, the final result for the beta functions which only
involves traces of products (and derivatives) of these matrices is
real. }.

We will leave the discussion of (untruncated) exact equations (and
to what extent we can find a justification for the truncations that
have been used) for future work. Here we will just follow what has
been done with the Wetterich equation with our version of the Polchinski
equation truncated to the lowest non-trivial operators for gravity
coupled to a scalar field. Then the beta function equations for the
dimension zero and two operators are given by (defining $\dot{x}=\Lambda\frac{d}{d\Lambda}x,\,\hat{\lambda}_{i}=\lambda_{i}/(Z_{0})^{i}\,\gamma_{i}=\ln Z_{i}$),
\begin{eqnarray}
\dot{g}_{0}+4g_{0} & = & \frac{1}{(4\pi)^{2}}[10e^{-g_{0}/g_{1}}-4+e^{-\hat{\lambda}_{1}}],\label{eq:g0beta}\\
\dot{g}_{1}+2g_{1} & = & -\frac{1}{(4\pi)^{2}}\frac{1}{3}[13e^{-g_{0}/g_{1}}+5+\frac{1}{2}e^{-\hat{\lambda}_{1}}(1-6\hat{\xi}_{1})],\label{eq:g1beta}\\
\dot{\hat{\lambda}}_{1}+\dot{\gamma}_{0}\hat{\lambda}_{1}+2\hat{\lambda}_{1} & = & -\frac{e^{-\hat{\lambda}_{1}}}{(4\pi)^{2}}[\frac{\hat{\lambda}_{2}}{2}+\frac{1}{8}\frac{\hat{\lambda}_{1}^{2}}{g_{1}}]-\frac{5}{(4\pi)^{2}}e^{-g_{0}/g_{1}}\frac{\hat{\lambda}_{1}}{g_{1}}.\label{eq:lam1beta}
\end{eqnarray}
Note that in the last equation we've kept the contribution of the
operator $\phi^{4}$ since this is a measure of the error introduced
by the truncation of the scalar field theory. It is useful also to
consider these equations in terms of the following alternative variables
\begin{equation}
g_{_{N}}(\Lambda)=2\kappa^{2}(\Lambda)\Lambda^{2}=-\frac{1}{g_{1}(\Lambda)},\,2\lambda_{CC}=\Lambda^{2}2\kappa^{2}g_{0}=-\frac{g_{0}}{g_{1}}.\label{eq:grelabel}
\end{equation}
The beta function equations above then become\footnote{Note that in previous derivations of these equations there is a singularity
$1/1-2\lambda_{CC})$. This is absent in our treatment and comes from
the exact form given below i.e. $e^{2\lambda_{CC}}=1/e^{-2\lambda_{CC}}$
which if expanded in the denominator and truncated to its leading
two terms gives the above singularity which is clearly spurious.}
\begin{eqnarray}
\dot{\lambda}_{{\rm CC}}+2\lambda_{{\rm CC}} & = & \frac{g_{N}}{(4\pi)^{2}}\left[(5-\frac{13}{3}\lambda_{{\rm CC}})e^{2\lambda_{{\rm CC}}}-(2+\frac{5}{3}\lambda_{{\rm CC}})+e^{-\hat{\lambda}_{1}}(\frac{1}{2}-\frac{1}{6}\lambda_{{\rm CC}}(1-6\hat{\xi}_{1}))\right],\label{eq:CCbeta}\\
\dot{g}_{N}-2g_{N} & = & -\frac{g_{N}^{2}}{(4\pi)^{2}}\frac{1}{3}\left[13e^{2\lambda_{{\rm CC}}}+5+\frac{1}{2}e^{-\hat{\lambda}_{1}}(1-6\hat{\xi}_{1})\right],\label{eq:gNbeta}\\
\dot{\hat{\lambda}}_{1}+\dot{\gamma}_{0}\hat{\lambda}_{1}+2\hat{\lambda}_{1} & = & -\frac{e^{-\hat{\lambda}_{1}}}{(4\pi)^{2}}[\frac{\hat{\lambda}_{2}}{2}-\frac{1}{8}g_{N}\hat{\lambda}_{1}^{2}]+\frac{5}{(4\pi)^{2}}e^{2\lambda_{CC}}g_{N}\hat{\lambda}_{1}.\label{eq:lam1beta2}
\end{eqnarray}
In this form we see the Gaussian fixed point (GFP) at $g_{N}=\lambda_{CC}=\lambda_{1}=\lambda_{2}=\xi_{1}=\xi_{2}=0$.
Clearly there is also a non-Gaussian fixed point (NGFP) with for instance
$g_{N}^{*}=6(4\pi)^{2}[13e^{2\lambda_{{\rm CC}}}+5+\frac{1}{2}e^{-\hat{\lambda}_{1}}(1-6\hat{\xi}_{1})]^{-1}$.
Note that this is clearly positive at least as long as $\xi_{1}$
is not $\gg O(1)$ as in Higgs inflation and that one may also get
(effectively) a large negative contribution if there is a large number
of scalar fields.

However to investigate the NGFP it is more transparent to consider
the previous version i.e. (\ref{eq:g0beta})(\ref{eq:g1beta})(\ref{eq:lam1beta}).
Putting the ``time'' derivatives to zero, we have a system of equations
which determine the fixed point vales of all the couplings in terms
of two undetermined parameters (both coming from scalar ``mass''
terms) which may be chosen to be (say) $\hat{\lambda}_{1}=\hat{\lambda}_{1}^{*},\,\hat{\xi}_{1}=\hat{\xi}_{1}^{*}$.
Then the first two fixed point equations (\ref{eq:g0beta})(\ref{eq:g1beta})
give one transcendental equation 
\[
2\lambda_{CC}=3\frac{10e^{2\lambda_{CC}}-4+e^{-\hat{\lambda}_{1}^{*}}}{26e^{2\lambda_{CC}}+10+e^{-\hat{\lambda}_{1}^{*}}(1-6\hat{\xi}_{1}^{*})}.
\]
This determines $\lambda_{CC}^{*}$ (as an $O(1)$ number) which when
used in the second equation will determine $g_{1}^{*}$ (which it
should be noted must be negative giving a positive value for $g_{N}$).
The third equation then determines the $\phi^{4}$ coupling $\hat{\lambda}_{2}^{*}$.
It also follows from the fact that the fixed point is at $O(1)$ values
for $g_{0},g_{1}$ (taking $\lambda_{1}^{*}>0$ and $\xi_{1}^{*}=O(1)$),
that the critical exponents are essentially given by the canonical
values up to corrections $O(1/(4\pi)^{2})$, so that this is indeed
a UV fixed point with (at least) two relevant directions. 

To get the beta function for $\lambda_{2}$ we need to consider the
heat kernel expansion to the next order i.e. the ${\bf b}_{4}$ term.
The coefficients of the beta function for the terms in the potential
(to arbitrary order) are actually easily evaluated since they can
be obtained by ignoring all derivative terms of $\phi$ as discussed
before. In particular the beta function for $\phi^{4}$ term is 
\begin{equation}
\frac{1}{4!}(\dot{\lambda}_{2}+\dot{\gamma}_{0}\hat{\lambda}_{2})=\frac{e^{-\hat{\lambda}_{1}}}{(4\pi)^{2}}(\frac{1}{8}\hat{\lambda}_{2}^{2}-\frac{1}{4!}\hat{\lambda}_{3}+\frac{1}{3}\frac{1}{g_{1}}\hat{\lambda}_{2}\hat{\lambda}_{1})+\frac{4g_{N}e^{2\lambda_{CC}}}{(4\pi)^{2}}\frac{1}{4!}\hat{\lambda}_{2}.\label{eq:lam2beta}
\end{equation}
At a fixed point we already know $\hat{\lambda}_{1},\hat{\lambda}_{2}$,
$\lambda_{CC},g_{N}$ (the last two and hence the last term on the
RHS above is of course known provided we ignore higher derivative
terms), so this equation just determines $\hat{\lambda}_{3}$. This
process obviously repeats itself ad infinitum with the beta function
equation at a fixed point for $\hat{\lambda}_{n}$ determining $\hat{\lambda}_{n+1}$,
since the set of fixed point equations for $\lambda_{m}$ with $m<n$
have already determined all the other couplings which enter into this
equation\footnote{This argument, for the case of flat space scalar field theory, seems
to have been first given by Weinberg \citep{Weinberg:1976xy}. For
comments on the usual statements about triviality of such theories
see also \citep{Weinberg:1996kr} page 137.}. 

Just focusing on the scalar sector in flat space it appears as if
there is a one parameter family of fixed points since $\lambda_{1}$
does not appear to be determined by any of the fixed point equations.
However it has been argued (see \citep{Morris:1998da}\citep{Codello:2012sc}
and references therein for flat space scalar field theory at least)
the requirement of a scaling solution for all values of the dimensionless
field $\tilde{\phi}\in\Re$ would restrict the allowed values of $\lambda_{1}$
to a discrete set for space time dimension $d=3$ (in the sense that
the potential blows up at some finite value of $\tilde{\phi}$ otherwise)
and has no solution in $d=4$. However it is not clear that there
is a physical requirement that the potential should be non-singular
for all real values of $\tilde{\phi}$. In practice for finite $\Lambda$
one works with $\tilde{\phi}\equiv\phi/\Lambda<\tilde{\phi_{s}}<1$
so it is not obvious that a singularity (in the limit $\Lambda\rightarrow\infty$)
at a finite value of $\tilde{\phi}_{s}$ is not physically reasonable.
If indeed the scalar field sector is trivial, then for the asymptotic
safety program to remain valid one needs to argue that the inclusion
of gravitational corrections (such as the last term of (\ref{eq:lam2beta})),
somehow changes this conclusion. 

The necessary inclusion of higher order terms such as $R^{2}$, implies
that we have higher derivative terms in the kinetic operator; i.e.
${\bf K}=-\nabla^{2}{\bf I}+{\bf E}+O(\nabla^{4})$. These terms will
not only affect ${\bf b}_{4}$ and the higher order coefficients but
also the lower order ones and hence the beta-functions for the cosmological
constant and Newton's constant as well. These calculations have been
performed for operators whose highest derivative is quartic by generalizing
standard heat kernel methods (see for example \citep{Gusynin:1988zt}\citep{Ohta:2013uca}).
Calculating these has been done for $f(R)$ theories and for a theory
with $R_{\mu\nu}^{2}+R^{2}$ (see for example \citep{Codello:2008vh}
and references therein). The general conclusion is that the fixed
point values obtained for the lowest order truncations (i.e. the solutions
of (\ref{eq:CCbeta})(\ref{eq:gNbeta}) etc.) are not changed significantly.

However in general even with the inclusion of all the higher order
contributions to ${\bf K}$, one expects that the coefficients of
the generalized heat kernel expansion will all be of $O(1)$ or smaller
as is the case for the ones that have actually been calculated, and
that there is no singular behavior in this expansion. If true this
implies that the anomalous dimension matrix has at most order one
coefficients divided by $(4\pi)^{2}$. This means that the relevance
or irrelevance of operators at the UV fixed point is determined by
their canonical dimensions - except for the three $R^{2}$ operators
and the $\phi^{4}$ interaction (see eqn (\ref{eq:deltagdot})) and
perhaps a few more. Thus we see that the cosmological constant term
and the Einstein Hilbert term are relevant\footnote{It is important to stress that these statements are manifest in terms
of the natural couplings $g_{i}$ defined in (\ref{eq:Igravscalar}).
If on the other hand one chose the inverses (such as $g_{N}=-g_{1}^{-1}$)
then the fixed point values are at numerically large values ($O(4\pi)^{2}$).
Thus (the equation (see (\ref{eq:g1beta}) and (\ref{eq:deltagdot}))
for $\delta g_{1}\equiv g_{1}-g_{1}^{*}$ is 
\[
\dot{\delta g}_{1}+2\delta g_{1}=O(\frac{1}{(4\pi)^{2}})
\]
 so that this is a relevant direction. On the other hand defining
$\delta g_{N}=g_{N}-g_{N}^{*}$ the equation gives $\dot{\delta g}_{N}-2\delta g_{N}\sim O(\frac{g_{N}}{(4\pi)^{2}})g_{N}$
which of course can be used to establish that near the GFP this direction
is relevant in the IR but near the NGFP which is strongly coupled
the right hand side cannot be ignored, and restores the conclusion
based on the analysis of the $g_{1}$ equation that direction is relevant
in the UV. See for example \citep{Codello:2008vh,Reuter:2012id}.} and all operators of dimension 6 and above are most probably irrelevant
since to make them relevant, one would need large anomalous dimensions
to cancel the canonical dimension term in (\ref{eq:deltagdot}) for
$n_{A}-4>2$. This is of course the general expectation that has been
confirmed in some cases by detailed calculations in the literature.
The relevance or irrelevance of the dimension four operators on the
other hand can only be established by direct calculation. Truncating
at this order, for instance in pure gravity, appears to give a non
trivial fixed point value for these couplings (see for example \citep{Percacci:2017}
and references therein). The relevance (or irrelevance) of these classically
marginal operators is however hard to establish since their scaling
dimension could be significantly affected by higher derivative ($"R^{6}"$
etc) terms. So it seems that the situation for these operators remains
somewhat murky.

\section{Comments and conclusions}

Some comments on the general approach we have adopted here and its
relation to the literature are in order.
\begin{enumerate}
\item The first equality of equation (\ref{eq:P3}) (or (\ref{eq:P4}))
resembles the Wetterich equation (\ref{eq:Wett}) if one uses $\frac{d}{dt}\ln K_{k,\Lambda}=\frac{\dot{K}_{k,\Lambda}}{K_{k,\Lambda}}=\frac{\dot{R}_{k,\Lambda}}{K_{k,\Lambda}}$
. This would then reintroduce the problem mentioned at the end of
section (\ref{sec:Review-and-some}). The point is to use the Schwinger
proper time expression (and regularization) of the one-loop determinant
given in (\ref{eq:regulator-1}). Note that the difference (in the
$k\rightarrow\Lambda$ limit) is an infinite term (which is not field
independent in the background field version which is essential for
the gauge theory discussion).
\item In practice, the average effective action is only used in the regime
where the background field momentum modes are low compared to the
cutoff $k$ (which is usually identified as a IR cutoff) in order
for the heat kernel expansion to be valid. So the entire discussion
involving the so-called average effective action is completely equivalent
to one involving the Wilsonian effective action. To put it another
way as far as deriving the beta function equations go, one is effectively
dealing with a floating UV cutoff rather than a IR cutoff. Although
it is formally true that the definition (\ref{eq:WettInteg}) leads
to the quantum effective action $\Gamma$ in the limit $k\rightarrow0$
it is difficult to recover in practice the non-localities of $\Gamma$
starting from the quasi-local $\Gamma_{k}$ since this would involve
summing the infinite series of the derivative expansion\footnote{For partial summations recovering some aspects of non-locality see
\citep{Codello:2015oqa}.}. 
\item For the same reason as in the above, the RG equation (\ref{eq:P4})
does not generate non-local terms - the expansion of the heat kernel
is local as long as the mode numbers (``momenta'') of the background
fields are small compared to the cutoff. This is of course the well
known statement that the Wilsonian effective action is (quasi) local.
By the above arguments it follows that in actual calculations the
average effective action is also being treated as being quasi-local. 
\item The use of the equation (\ref{eq:P4}) coupled with the heat kernel
expansion and proper time regularization gives a simple expression
for all the beta function equations of the theory. Of course as in
other approaches one can find the fixed point only after truncating
the number of operators. 
\item To the extent that they have been computed, the heat kernel coefficients
are order one or smaller. Given that there is no reason to expect
anomalously large coefficients it is very plausible that there is
only a finite number of relevant directions at the UV fixed point.
In other words we expect that there are no large anomalous dimensions
which can swamp the canonical dimensions of operators, which grow
with the order of the terms in the derivative expansion. We found
two relevant directions in agreement with previous work in pure gravity. 
\item While in pure gravity the UV fixed point is completely determined,
in gravity coupled to a scalar field theory (or to the standard model),
there are two undetermined parameters corresponding to the scalar
mass term and the $\phi^{2}R$ term. However as in the case of flat
space scalar field theory, it is possible that only a discrete set
of values admit a scaling solution . All other couplings are then
determined in terms of these. 
\end{enumerate}
Much work remains to be done to really establish the standard model
coupled to gravity as a UV complete theory. While an iterative procedure
for establishing the existence of a fixed point appears to be valid
for the scalar potential the same is not the case for derivative interactions
since higher derivative terms feed back into the flow equation for
lower derivative terms. Nevertheless explicit calculations \citep{Codello:2008vh}\citep{Reuter:2007rv}
seem to show that the fixed point established (in pure gravity) is
hardly affected by the inclusion of higher derivative terms. Also
apart from giving a rigorous proof of the dimensionality of the critical
surface, an important unresolved problem remains; that of establishing
the unitarity of the theory, since in perturbation theory such higher
derivative theories have ghosts. As discussed in subsection (\ref{sub:Gauge-fixing-dependence})
at tree level these ghosts will appear at the (low energy) Planck
mass. However at the scale of this putative ghost, in order to minimize
radiative corrections in the Wilsonian action, the cutoff $\Lambda$
needs to be chosen at around the same scale. But in this case all
terms in the effective action will make similar contributions. On
the other hand if the cutoff is chosen much larger the radiative corrections
become important. Thus it is possible that (as discussed also in section
VIII of \citep{Codello:2008vh} for instance) non-perturbative quantum
effects change the spectrum at the putative ghost mass. It is also
possible that higher derivative theories need to be reinterpreted
as having PT invariant rather than Hermitian Hamiltonians as suggested
in \citep{Bender:2007wu,Bender:2008vh,Mannheim:2018ljq}. Clearly
this is one of the most important issue that needs to be addressed
in the asymptotic safety program.

\section*{Acknowledgements}

I wish to thank Joe Polchinski for useful comments on the manuscript
and emphasizing the unresolved issue of unitarity. Thanks also to
Cliff Burgess and especially to Leo Pando Zayas for reading the manuscript
and for asking several questions which led me to explain some of the
arguments in greater detail. I also wish to thank Astrid Eichhorn
for a discussion on various issues related to the asymptotic safety
program and for drawing my attention to several references. Finally
many thanks to Roberto Percacci for carefully reading the manuscript,
several discussions on the asymptotic safety program and for many
references and valuable suggestions. I also wish to acknowledge the
hospitality of the Abdus Salam ICTP and its director Fernando Quevedo
for hospitality during the completion of this work.

\bibliographystyle{apsrev}
\bibliography{myrefs}

\end{document}